\documentclass[twocolumn,notitlepage,nofootinbib]{revtex4-1}
\usepackage{amsmath,amssymb,bm,graphicx,hyperref}
\allowdisplaybreaks[1]

\renewcommand\d{\partial}
\newcommand\+{\dagger}
\newcommand\<{\langle}
\renewcommand\>{\rangle}
\newcommand\eps{\varepsilon}
\newcommand\0{{\bm{0}}}
\renewcommand\k{{\bm{k}}}
\newcommand\p{{\bm{p}}}
\newcommand\q{{\bm{q}}}
\renewcommand\r{{\bm{r}}}
\renewcommand\v{{\bm{v}}}
\newcommand\C{\mathcal{C}}
\newcommand\E{\mathcal{E}}
\renewcommand\H{\mathcal{H}}
\newcommand\J{\mathcal{J}}
\newcommand\N{\mathcal{N}}
\renewcommand\O{\mathcal{O}}
\newcommand\Q{\mathcal{Q}}
\renewcommand\S{\mathcal{S}}
\newcommand\X{\mathcal{X}}
\newcommand\Y{\mathcal{Y}}
\DeclareMathOperator\Tr{Tr}
\let\Re\relax\DeclareMathOperator\Re{Re}
\let\Im\relax\DeclareMathOperator\Im{Im}

\begin{document}

\title{Bulk viscosity of resonating fermions revisited:\\
Kubo formula, sum rule, and the dimer and high-temperature limits}

\author{Keisuke Fujii}
\author{Yusuke Nishida}
\affiliation{Department of Physics, Tokyo Institute of Technology,
Ookayama, Meguro, Tokyo 152-8551, Japan}

\date{April 2020}

\begin{abstract}
The bulk viscosity of two-component fermions with a zero-range interaction is revisited both in two and three dimensions.
We first point out that the ``standard'' Kubo formula employed in recent studies has flaws to give rise to an unphysical divergent bulk viscosity even in a limit where it is supposed to vanish.
The corrected Kubo formula as well as the sum rule is then carefully rederived so as to confirm that the bulk viscosity indeed vanishes in the free, unitarity, and dimer limits.
We also discuss that the recently found discrepancy between the Kubo formalism and the kinetic theory for the bulk viscosity is attributed to the fact that the quasiparticle approximation assumed by the latter breaks down even in the high-temperature limit.
\end{abstract}

\maketitle

\section{Introduction}\label{sec:intro}
Two-component fermions with a zero-range interaction constitute a system of simple elegance that is parametrized solely by a scattering length, $a$~\cite{Zwerger:2012}.
As the inverse scattering length increases, the system evolves from a free Fermi gas (free limit) to a free Bose gas of tightly bound dimers (dimer limit).%
\footnote{The free and dimer limits are often referred to as the BCS (Bardeen-Cooper-Schrieffer) and BEC (Bose-Einstein condensation) limits, respectively, which are however avoided in this paper because we do not necessarily work below the superfluid critical temperature.}
In particular, when the scattering length diverges (unitarity limit), the scale and conformal invariance emerge~\cite{Mehen:2000,Son:2006,Nishida:2007}, so that its equation of state obeys the ideal gas law although the system is strongly interacting.
The conformal invariance manifests itself also in dynamic properties such as the vanishing bulk viscosity~\cite{Son:2007,Taylor:2010,Enss:2011}.

Recently, the frequency-dependent bulk viscosity (bulk viscosity spectral function) for an arbitrary scattering length was studied both in two and three dimensions based on the quantum virial expansion~\cite{Nishida:2019,Enss:2019,Hofmann:2020}.
The starting point was the ``standard'' Kubo formula for the bulk viscosity,
\begin{align}\label{eq:kubo_stress}
\Re[\zeta(\omega)] = \frac{\Im[R_{\Pi\Pi}(\omega)]}{\omega},
\end{align}
where
\begin{align}
R_{XY}(\omega) = \frac{i}{L^d}\int_0^\infty\!dt\,e^{i(\omega+i0^+)t}
\<[\hat{X}(t),\hat{Y}(0)]\>
\end{align}
for $\hat{X}=\hat{Y}=\hat\Pi$ is the stress-stress response function at zero wave vector.
Because the trace of the integrated stress tensor operator is provided by $d\cdot\hat\Pi=2\hat{H}+\hat{C}/(\Omega_{d-1}ma^{d-2})$ and the commutator of the Hamiltonian with any operator in the grand canonical average vanishes, the above Kubo formula turns into the favorite form of
\begin{align}\label{eq:kubo_contact}
\Re[\zeta(\omega)] = \frac1{(d\cdot\Omega_{d-1}ma^{d-2})^2}\frac{\Im[R_{CC}(\omega)]}{\omega},
\end{align}
where $\hat{C}$ is the contact operator~\cite{Martinez:2017,Fujii:2018}.
It is its zero-frequency limit that corresponds to the bulk viscosity in hydrodynamics.
The latter formula was then evaluated systematically in the high-temperature limit where the fugacity serves as a small expansion parameter~\cite{Nishida:2019,Enss:2019,Hofmann:2020}.

Actually, these formulas have both technical and physical flaws (see also Ref.~\cite{Bradlyn:2012}).
In order to derive Eq.~(\ref{eq:kubo_contact}), three terms such as $\Im[R_{HH}(\omega)]/\omega$ are dropped in Eq.~(\ref{eq:kubo_stress}) on the ground that the numerator vanishes.
However, caution is required in the zero-frequency limit because the denominator also vanishes.
Indeed, by employing the spectral representation,
\begin{align}
R_{XY}(\omega)
&= -\frac1{L^dZ}\sum_{m,n}\frac{e^{-\beta E_m}-e^{-\beta E_n}}{\omega+E_m-E_n+i0^+} \notag\\
&\quad \times \<m|\hat{X}|n\>\<n|\hat{Y}|m\>,
\end{align}
and taking its imaginary part, one finds
\begin{align}
\frac{\Im[R_{HH}(\omega)]}{\omega}
= \frac\pi{L^dZ}\,\delta(\omega)\sum_n\beta e^{-\beta E_n}E_n^2
\end{align}
for $\hat{X}=\hat{Y}=\hat{H}$.
This term thus diverges at zero frequency for an arbitrary scattering length including the free and unitarity limits where the bulk viscosity is supposed to vanish.
Whether this and the other two terms should be dropped or not is ambiguous if one starts with Eq.~(\ref{eq:kubo_stress}).

Even if one takes Eq.~(\ref{eq:kubo_contact}) for granted, which now vanishes in the free and unitarity limits, it gives rise to a term proportional to $\delta(\omega)/a^4$ in the dimer limit (see Footnote \ref{footnote} at the end of Sec.~\ref{sec:dimer}).
Because the system in the dimer limit is a free Bose gas of tightly bound dimers, it should exhibit scale invariance if probed at a lower frequency than their binding energy.
Therefore, our physical intuition supposes that the bulk viscosity vanishes again, which conflicts with the divergent bulk viscosity of Eq.~(\ref{eq:kubo_contact}) in the dimer limit.

The purpose of this paper is to demonstrate that the above flaws are resolved by correcting the Kubo formula in Eq.~(\ref{eq:kubo_stress}).
Although the corrected Kubo formula has been known since long ago~\cite{Mori:1962,Luttinger:1964}, it is not well appreciated by the literature in the context of ultracold atom physics.
Therefore, we first review its derivation as well as the sum rule in Sec.~\ref{sec:kubo} and then carefully evaluate the corrected Kubo formula in Sec.~\ref{sec:limits}, confirming that the bulk viscosity indeed vanishes in the free, unitarity, and dimer limits.
We also revisit the bulk viscosity in the high-temperature limit in Sec.~\ref{sec:high-T} and discuss a possible origin of the discrepancy between the Kubo formalism and the kinetic theory found recently in Refs.~\cite{Nishida:2019,Enss:2019,Hofmann:2020}.
Finally, Sec.~\ref{sec:summary} is devoted to a summary of this paper and some useful formulas regarding Kubo's canonical correlation function are presented in Appendix~\ref{sec:canonical}.

In what follows, we will set $\hbar=k_B=1$ and implicit summations over repeated indices are assumed throughout this paper.
Also, an integration over $d$-dimensional wave vector or momentum is denoted by $\int_\k\equiv\int\!d\k/(2\pi)^d$ for the sake of brevity.

\section{Kubo formula}\label{sec:kubo}
The Kubo formula for the bulk viscosity can be derived by matching current responses against an external force between microscopic and low-energy effective descriptions, the latter of which is of course hydrodynamics.
Our derivation reviewed in this section partly follows that in Ref.~\cite{Luttinger:1964} (see Appendix~B therein).

\subsection{Microscopics}
We first consider that the system is weakly perturbed by an external vector potential, so that the microscopic Hamiltonian reads
\begin{align}
\hat{H} \to \hat{H}^A(t)
= \int\!d\r\,\frac{[D_i\hat\psi_\sigma(\r)]^\+[D_i\hat\psi_\sigma(\r)]}{2m} + \hat{V},
\end{align}
where $m$ is a mass of particles and $D_i\equiv\d_i-iA_i(t,\r)$ is the covariant derivative.
Accordingly, the current density operator is modified into
\begin{align}\label{eq:current}
\hat\J_i^A(t,\r) \equiv -\frac{\delta\hat{H}^A(t)}{\delta A_i(t,\r)}
= \hat\J_i(\r) - \hat\N(\r)\frac{A_i(t,\r)}{m}
\end{align}
with $\hat\N(\r)=\hat\psi_\sigma^\+(\r)\hat\psi_\sigma(\r)$ and $\hat\J_i(\r)=[\hat\psi_\sigma^\+(\r)\d_i\hat\psi_\sigma(\r)-\d_i\hat\psi_\sigma^\+(\r)\hat\psi_\sigma(\r)]/(2im)$ being the unperturbed number and current density operators, respectively.
The linear-response theory predicts that the expectation value of Eq.~(\ref{eq:current}) is provided by
\begin{align}
\J_i(t,\r) &= \<\hat\J_i^A(\r)\>
+ i\int_{-\infty}^t\!dt'\int\!d\r'\<[\hat\J_i(t,\r),\hat\J_j(t',\r')]\> \notag\\
&\quad \times A_j(t',\r') + O(A^2),
\end{align}
where $\hat\O(t,\r)\equiv e^{i\hat{H}t}\hat\O(\r)e^{-i\hat{H}t}$ is an operator in the Heisenberg representation and $\<\cdots\>\equiv\Tr[e^{-\beta(\hat{H}-\mu\hat{N})}\cdots]/\Tr[e^{-\beta(\hat{H}-\mu\hat{N})}]$ is an expectation value without the perturbation~\cite{Altland-Simons}.
Then, by setting $\<\hat\N(\r)\>=\N$ and $\<\hat\J_i(\r)\>=0$ in thermodynamic equilibrium, the spacetime Fourier transformation leads to
\begin{align}
\J_i(w,\k) = -\N\frac{A_i(w,\k)}{m} + R_{\J_i\J_j}(w,\k)A_j(w,\k) + O(A^2),
\end{align}
where
\begin{align}\label{eq:response}
R_{\X\Y}(w,\k) \equiv i\int_0^\infty\!dt\int\!d\r\,e^{iwt-i\k\cdot\r}
\<[\hat\X(t,\r),\hat\Y(0,\0)]\>
\end{align}
is a response function and $w$ denotes an arbitrary complex frequency with $\Im[w]>0$.
Although $w$ is eventually replaced by $\omega+i0^+$ for a real frequency $\omega$, it is of technical help to work on the upper-half plane of complex $w$ until the very end of all calculations.

It will turn out to be favorable to express the current-current response function in terms of Kubo's canonical correlation function,
\begin{align}\label{eq:canonical}
K_{\X\Y}(w,\k) &\equiv \int_0^\infty\!dt\int\!d\r\,e^{iwt-i\k\cdot\r}
\int_0^\beta\!\frac{d\tau}{\beta} \notag\\
&\quad \times \<\delta\hat\X(t-i\tau,\r)\delta\hat\Y(0,\0)\>,
\end{align}
where $\delta\hat\O(t,\r)\equiv\hat\O(t,\r)-\<\hat\O(t,\r)\>$ is an operator with its expectation value subtracted~\cite{Kubo:1957}.
After some calculations as detailed in Appendix~\ref{sec:current-stress}, we obtain
\begin{align}\label{eq:current-stress}
iwR_{\J_i\J_j}(w,\k) = -\beta K_{\pi_{ik}\pi_{jl}}(w,\k)\frac{k_kk_l}{m^2},
\end{align}
where $\hat\pi_{ij}(\r)$ is the unperturbed stress tensor operator obeying the momentum continuity equation,
\begin{align}\label{eq:continuity}
m\d_t\hat\J_i(t,\r) + \d_j\hat\pi_{ij}(t,\r) = 0.
\end{align}
Therefore, the current response is found to be
\begin{align}\label{eq:microscopic}
\J_i(w,\k) &= -\left[m\N\delta_{ij}
+ \beta K_{\pi_{ik}\pi_{jl}}(w,\k)\frac{k_kk_l}{iw}\right]\frac{A_j(w,\k)}{m^2} \notag\\
&\quad + O(A^2)
\end{align}
in the microscopic description.

\subsection{Hydrodynamics}
We then consider that the system perturbed at low frequency and wave vector is described by hydrodynamics, which is founded on the number continuity equation,
\begin{align}\label{eq:number}
\d_t\N(t,\r) + \d_i\J_i(t,\r) = 0,
\end{align}
the momentum continuity equation,
\begin{align}
& m\d_t\J_i(t,\r) + \d_j\pi_{ij}(t,\r) \notag\\
&= \N(t,\r)E_i(t,\r) + \J_j(t,\r)F_{ij}(t,\r),
\end{align}
and the energy continuity equation,
\begin{align}\label{eq:energy}
\d_t\H(t,\r) + \d_i\Q_i(t,\r) = \J_i(t,\r)E_i(t,\r).
\end{align}
Here, $E_i(t,\r)=-\d_tA_i(t,\r)$ and $F_{ij}=\d_iA_j(t,\r)-\d_jA_i(t,\r)$ are the external electric and magnetic fields, respectively, and the conserved charge densities and their fluxes are to be expressed in terms of the local thermodynamic variables and the fluid flow velocity $v_i(t,\r)$.
The constitutive relations for normal fluids read
\begin{align}
\J_i(t,\r) = \N(t,\r)v_i(t,\r)
\end{align}
for the number current density,
\begin{align}
\H(t,\r) = \E(t,\r) + \frac{m}{2}\N(t,\r)[\v(t,\r)]^2
\end{align}
for the energy density,
\begin{align}
\pi_{ij}(t,\r) &= p(t,\r)\delta_{ij} + m\N(t,\r)v_i(t,\r)v_j(t,\r) \notag\\
&\quad - \sigma_{ij}(t,\r)
\end{align}
for the stress tensor,
\begin{align}
\Q_i(t,\r) &= [\H(t,\r)+p(t,\r)]v_i(t,\r) \notag\\
&\quad - \sigma_{ij}(t,\r)v_j(t,\r) - \kappa\,\d_iT(t,\r)
\end{align}
for the energy current density with
\begin{align}
\sigma_{ij} &= \left(\zeta-\frac2d\,\eta\right)\delta_{ij}\d_kv_k(t,\r) \notag\\
&\quad + \eta\,[\d_iv_j(t,\r)+\d_jv_i(t,\r)],
\end{align}
where $\zeta$ is the bulk viscosity, $\eta$ is the shear viscosity, and $\kappa$ is the thermal conductivity~\cite{Landau-Lifshitz}.
We choose the number density $\N(t,\r)$ and the internal energy density $\E(t,\r)$ as the independent variables, so that the pressure $p(t,\r)=p[\N(t,\r),\E(t,\r)]$ and the temperature $T(t,\r)=T[\N(t,\r),\E(t,\r)]$ are locally determined by the equations of state.

When the perturbation by the external vector potential is weak, the thermodynamic variables slightly deviate from their equilibrium values, so that $\delta\N(t,\r)=\N(t,\r)-\N$, $\delta\E(t,\r)=\E(t,\r)-\E$, and $v_i(t,\r)$ are as small as $O(A)$.
After linearizing the hydrodynamic equations in Eqs.~(\ref{eq:number})--(\ref{eq:energy}), the spacetime Fourier transformation leads to
\begin{align}
& -iw\delta\N(w,\k) + \N ik_iv_i(w,\k) = 0, \\[2pt]
& -m\N iwv_i(w,\k) + \left(\frac{\d p}{\d\N}\right)_\E ik_i\delta\N(w,\k) \notag\\
&\quad + \left(\frac{\d p}{\d\E}\right)_\N ik_i\delta\E(w,\k)
+ \left(\zeta+\frac{d-2}{d}\eta\right)k_ik_jv_j(w,\k) \notag\\
&\quad + \eta\,\k^2v_i(w,\k) = \N iwA_i(w,\k), \\[8pt]
& -iw\delta\E(w,\k) + (\E+p)ik_iv_i(w,\k) \notag\\
&\quad + \kappa\left(\frac{\d T}{\d\N}\right)_\E\k^2\delta\N(w,\k)
+ \kappa\left(\frac{\d T}{\d\E}\right)_\N\k^2\delta\E(w,\k) = 0.
\end{align}
Finally, by eliminating $\delta\N(w,\k)$ and $\delta\E(w,\k)$, the current response up to $O(k^2)$ is found to be
\begin{align}\label{eq:hydrodynamic}
& \J_i(w,\k) = -\,\biggl[m\N\delta_{ij}
- \left(\frac{\d p}{\d\N}\right)_\E\N\frac{k_ik_j}{(iw)^2} \notag\\
&\quad - \left(\frac{\d p}{\d\E}\right)_\N(\E+p)\frac{k_ik_j}{(iw)^2}
+ \left(\zeta+\frac{d-2}{d}\eta\right)\frac{k_ik_j}{iw} \notag\\
&\quad + \eta\,\frac{\k^2}{iw}\delta_{ij} + O(k^3)\biggr]\,\frac{A_j(w,\k)}{m^2} + O(A^2)
\end{align}
in the hydrodynamic description.

Here, it is worthwhile to emphasize that the second and third terms in the square brackets of Eq.~(\ref{eq:hydrodynamic}) originate from the pressure fluctuations associated with the fluctuations of the number and energy densities, respectively, which are essential to the correct Kubo formula for the bulk viscosity~\cite{Mori:1962}.
However, such pressure fluctuations were neglected in Ref.~\cite{Taylor:2010} by stating \textit{``In the long-wavelength limit, the contributions to the stress tensor coming from viscous terms dominate over contributions from pressure fluctuations,''} which we find ungrounded because both the contributions are $O(k^2)$.
We also note that the second and third terms are combined into the sound velocity,
\begin{align}\label{eq:sound}
mc_s^2 \equiv \left(\frac{\d p}{\d\N}\right)_{\S/\N}
= \left(\frac{\d p}{\d\N}\right)_\E + \left(\frac{\d p}{\d\E}\right)_\N\frac{\E+p}{\N},
\end{align}
so as to relate the pressure fluctuations to the gapless sound mode with $\S$ being the entropy density.

\subsection{Bulk viscosity}
Now, by matching the current responses between the microscopic and hydrodynamic descriptions in Eqs.~(\ref{eq:microscopic}) and (\ref{eq:hydrodynamic}) at low frequency and wave vector, we obtain
\begin{align}
& \lim_{w\to i0^+}\left[\beta K_{\pi_{ik}\pi_{jl}}(w,\0)\frac{k_kk_l}{\k^2}
+ \frac{mc_s^2\N}{iw}\frac{k_ik_j}{\k^2}\right] \notag\\
&= \left(\zeta+\frac{d-2}{d}\eta\right)\frac{k_ik_j}{\k^2} + \eta\,\delta_{ij}.
\end{align}
Here, $K_{\pi_{ik}\pi_{jl}}(w,\0)$ is symmetric under the exchanges of $i\leftrightarrow k$ and $j\leftrightarrow l$ by definition of the stress tensor operator as well as under $(ik)\leftrightarrow(jl)$ according to the Onsager reciprocal relations.
Because the rotational invariance dictates that such a fourth-order tensor is decomposed into a sum of $\delta_{ik}\delta_{jl}$ and $\delta_{ij}\delta_{kl}+\delta_{il}\delta_{jk}$, we find
\begin{align}
& \lim_{w\to i0^+}\left[\beta K_{\pi_{ik}\pi_{jl}}(w,\0)
+ \frac{mc_s^2\N}{iw}\delta_{ik}\delta_{jl}\right] \notag\\
&= \left(\zeta-\frac2d\,\eta\right)\delta_{ik}\delta_{jl}
+ \eta\,(\delta_{ij}\delta_{kl}+\delta_{il}\delta_{jk}),
\end{align}
so that the bulk and shear viscosities are provided by
\begin{align}\label{eq:bulk_static}
\zeta &= \lim_{w\to i0^+}\left[\beta K_{\pi\pi}(w,\0) + \frac{mc_s^2\N}{iw}\right], \\
\eta &= \lim_{w\to i0^+}\beta K_{\pi_{xy}\pi_{xy}}(w,\0),
\end{align}
where $d\cdot\hat\pi(\r)\equiv\hat\pi_{ii}(\r)$ is the trace of the stress tensor operator.

It is customary to refer to the right-hand side of Eq.~(\ref{eq:bulk_static}),
\begin{align}\label{eq:bulk_dynamic}
\zeta(w) \equiv \beta K_{\pi\pi}(w,\0) + \frac{mc_s^2\N}{iw},
\end{align}
as a frequency-dependent complex bulk viscosity for $w\to\omega+i0^+$.
Because the bulk viscosity is provided by $\zeta=\lim_{w\to i0^+}\zeta(w)$, the singularity of the second term at $w=0$ originating from the gapless sound mode should be canceled by the same singularity inherent in the first term.
Actually, the two terms can elegantly be combined so as to modify the stress tensor operator as
\begin{align}\label{eq:fluctuation}
\hat{\tilde\pi}(\r) \equiv \hat\pi(\r)
- \left[\left(\frac{\d p}{\d\N}\right)_\E\hat\N(\r)
+ \left(\frac{\d p}{\d\E}\right)_\N\hat\H(\r)\right],
\end{align}
where the subtracted terms represent the pressure fluctuations with $\hat\N(\r)$ and $\hat\H(\r)$ being the number and energy density operators, respectively.
After some calculations as detailed in Appendix~\ref{sec:fluctuation}, we obtain the succinct form of
\begin{align}\label{eq:bulk_kubo}
\zeta(w) = \beta K_{\tilde\pi\tilde\pi}(w,\0),
\end{align}
which is nothing other than the Kubo formula for the bulk viscosity~\cite{Mori:1962,Luttinger:1964}.
We note that the canonical correlation function is favorable to clean up the rather involved expression in terms of the stress-stress response function~\cite{Bradlyn:2012}, as detailed in Appendix~\ref{sec:comparison}.

\subsection{Sum rule}
The sum rule for the frequency-dependent complex bulk viscosity from Eq.~(\ref{eq:bulk_dynamic}) reads
\begin{align}\label{eq:sum-rule}
& \int_{-\infty}^\infty\!\frac{d\omega}{\pi}\,\zeta(\omega+i0^+) \notag\\
&= \int\!d\r\int_0^\beta\!d\tau\,\<\delta\hat\pi(-i\tau,\r)\delta\hat\pi(0,\0)\>
- mc_s^2\N,
\end{align}
where the frequency integration sets the two operators at equal time.
In order to further evaluate it, we from now on specialize to two-component fermions with a zero-range interaction in $d$ spatial dimensions, for which the trace of the stress tensor operator is provided by
\begin{align}\label{eq:stress}
\hat\pi(\r) = \frac{2\hat\H(\r)}{d} + \frac{\hat\C(\r)}{d\cdot\Omega_{d-1}ma^{d-2}}
\end{align}
up to irrelevant total derivatives~\cite{Fujii:2018}.
Here, $\Omega_{d-1}\equiv(4\pi)^{d/2}/2\Gamma(2-d/2)=2,\,2\pi,\,4\pi$ coincides with the surface area of the unit $(d-1)$-sphere for $d=1,\,2,\,3$ and $\hat\C(\r)$ is the contact density operator~\cite{Braaten:2008}, which is related to the derivative of the Hamiltonian density with respect to the scattering length as
\begin{align}
\hat\C(\r) = \Omega_{d-1}ma^{d-1}\frac{\d\hat\H(\r)}{\d a}.
\end{align}
Accordingly, the derivative of the stress tensor operator with respect to the scattering length turns into
\begin{align}
\frac{\d\hat\pi(\r)}{\d a} = \frac{(4-d)\,\hat\C(\r)}{d\cdot\Omega_{d-1}ma^{d-1}},
\end{align}
because of $\d\hat\C(\r)/\d a=0$ for $2\leq d<4$.%
\footnote{This follows from $\hat\C(\r)\equiv(mg)^2\hat\psi_\sigma^\+(\r)\hat\psi_\tau^\+(\r)\hat\psi_\tau(\r)\hat\psi_\sigma(\r)/2$, $mg=\Omega_{d-1}(d-2)/[a^{2-d}-\Lambda^{d-2}/\Gamma(d/2)\Gamma(2-d/2)]$, $\d\hat\C(\r)/\d a=2mg\hat\C(\r)/(\Omega_{d-1}a^{d-1})$, and $g\to0$ in the limit of $\Lambda\to\infty$ [see also Eq.~(\ref{eq:coupling}) below].}
The spatial integrals of $\hat\N(\r)$, $\hat\H(\r)$, $\hat\C(\r)$, and $\hat\pi(\r)$ are to be denoted by $\hat{N}$, $\hat{H}$, $\hat{C}$, and $\hat\Pi$, respectively, and their expectation values by $\O=\<\hat\O(\r)\>$ except for the pressure $p=\<\hat\pi(\r)\>$.

Then, by employing the following properties of the canonical correlation function at equal time,%
\footnote{Here, it is helpful to recall $\d e^{-\beta\hat{H}}/\d a=-\int_0^\beta\!d\tau\,e^{-(\beta-\tau)\hat{H}}(\d\hat{H}/\d a)e^{-\tau\hat{H}}$, which follows from $h(\beta)\equiv e^{\beta\hat{H}}\d e^{-\beta\hat{H}}/\d a=\int_0^\beta\!d\tau\,h'(\tau)$ and $h'(\tau)=-e^{\tau\hat{H}}(\d\hat{H}/\d a)e^{-\tau\hat{H}}$.}
\begin{align}
& \int_0^\beta\!d\tau\,\<\delta\hat{H}(-i\tau)\delta\hat\O(\0)\>
= -\beta\left(\frac{\d\<\hat\O(\0)\>}{\d\beta}\right)_{\beta\mu,a}, \\
& \int_0^\beta\!d\tau\,\<\delta\hat{C}(-i\tau)\delta\hat\O(\0)\> \notag\\
&= -\Omega_{d-1}ma^{d-1}\left[\left(\frac{\d\<\hat\O(\0)\>}{\d a}\right)_{\beta,\mu}
- \left\<\frac{\d\hat\O(\0)}{\d a}\right\>\right],
\end{align}
as well as the thermodynamic identities,%
\footnote{They follow from the generalized Gibbs-Duhem relation, $dp=\S dT+\N d\mu-\C da/(\Omega_{d-1}ma^{d-1})$, including the differential of the scattering length~\cite{Tan:2008a,Tan:2008b,Tan:2008c}, and the Euler relation, $p=T\S+\mu\N-\E$.}
\begin{align}
\left(\frac{\d p}{\d\beta}\right)_{\beta\mu,a} &= -\frac{\E+p}{\beta}, \\
\left(\frac{\d p}{\d a}\right)_{\beta,\mu} &= -\frac{\C}{\Omega_{d-1}ma^{d-1}},
\end{align}
the sum rule for the frequency-dependent complex bulk viscosity is found to be
\begin{align}
\int_{-\infty}^\infty\!\frac{d\omega}{\pi}\,\zeta(\omega+i0^+)
= \frac{d+2}{d}p + \frac{(4-d)\,\C}{d^2\cdot\Omega_{d-1}ma^{d-2}} - mc_s^2\N.
\end{align}
Our sum rule determined solely by thermodynamics turns out to coincide with that derived in Ref.~\cite{Taylor:2010}, which we find unexpected because the last term originates from the pressure fluctuations neglected therein.
Finally, the thermodynamic identities together with the dimensional analysis, as detailed in Appendix~E of Ref.~\cite{Taylor:2010},%
\footnote{The dimensional analysis dictates $p=\N^{(d+2)/d}\,\tilde p(\N^{1/d}a,\S/\N)$ and thus $mc_s^2\N=(d+2)p/d+(a/d)(\d p/\d a)_{\N,\S}$, where the last term is further evaluated with Tan's pressure relation, $p=2\E/d+\C/(d\cdot\Omega_{d-1}ma^{d-2})$, and adiabatic relation, $(\d\E/\d a)_{\N,\S}=\C/(\Omega_{d-1}ma^{d-1})$~\cite{Tan:2008a,Tan:2008b,Tan:2008c}.}
simplifies the sum rule into
\begin{align}\label{eq:sum-rule_zero-range}
\int_{-\infty}^\infty\!\frac{d\omega}{\pi}\,\zeta(\omega+i0^+)
= -\frac{a^{3-d}}{d^2\cdot\Omega_{d-1}m}\left(\frac{\d\C}{\d a}\right)_{\N,\S}.
\end{align}
Here, both $\N$ and $\S$ (as opposed to $\S/\N$~\cite{Taylor:2010,Taylor:2012}) should be fixed in differentiating $\C$ with respect to $a$.

\section{Free, unitarity, and dimer limits}\label{sec:limits}
We evaluate the Kubo formula for the frequently-dependent complex bulk viscosity derived in the previous section, whose real part is supposed to vanish at an arbitrary frequency in the free and unitarity limits and at a lower frequency than the binding energy of dimers in the dimer limit.

\subsection{Free and unitarity limits}
In the free and unitarity limits where the system is scale invariant, the last term of the stress tensor operator in Eq.~(\ref{eq:stress}) is negligible because $\hat\C(\r)$ vanishes in the free limit and $a$ diverges in the unitarity limit.
Accordingly, the equation of state obeys the ideal gas law, $p=2\E/d$, so that the modified stress tensor operator in Eq.~(\ref{eq:fluctuation}) reads $\hat{\tilde\pi}(\r)=0$.
Therefore, the frequency-dependent complex bulk viscosity is found to vanish at an arbitrary frequency,
\begin{align}
\zeta(\omega+i0^+) = 0,
\end{align}
without any ambiguity discussed in Sec.~\ref{sec:intro} because the operator evaluated by the Kubo formula in Eq.~(\ref{eq:bulk_kubo}) is identically zero.

\subsection{Contact correlation}\label{sec:contact}
Although the canonical correlation function provides the succinct form of the frequently-dependent complex bulk viscosity, the response function is of practical help because the standard diagrammatic method can be applied.
As detailed in Appendix~\ref{sec:comparison}, Eq.~(\ref{eq:bulk_dynamic}) can be expressed in terms of the stress-stress response function and the sum rule as
\begin{align}
\zeta(w) = \frac{R_{\pi\pi}(w,\0)}{iw}
- \frac1{iw}\int_{-\infty}^\infty\!\frac{d\omega}{\pi}\,\zeta(\omega+i0^+).
\end{align}
In particular, for two-component fermions with a zero-range interaction, the substitution of Eqs.~(\ref{eq:stress}) and (\ref{eq:sum-rule_zero-range}) leads to
\begin{align}\label{eq:bulk_zero-range}
\zeta(w) = \frac1{iw}\frac{R_{\C\C}(w,\0)}{(d\cdot\Omega_{d-1}ma^{d-2})^2}
+ \frac1{iw}\frac{a^{3-d}}{d^2\cdot\Omega_{d-1}m}\left(\frac{\d\C}{\d a}\right)_{\N,\S},
\end{align}
where the commutator of the Hamiltonian with any operator in the grand canonical average can safely be dropped by working on the upper-half plane of complex $w$.

\begin{figure}[t]
\includegraphics[width=\columnwidth]{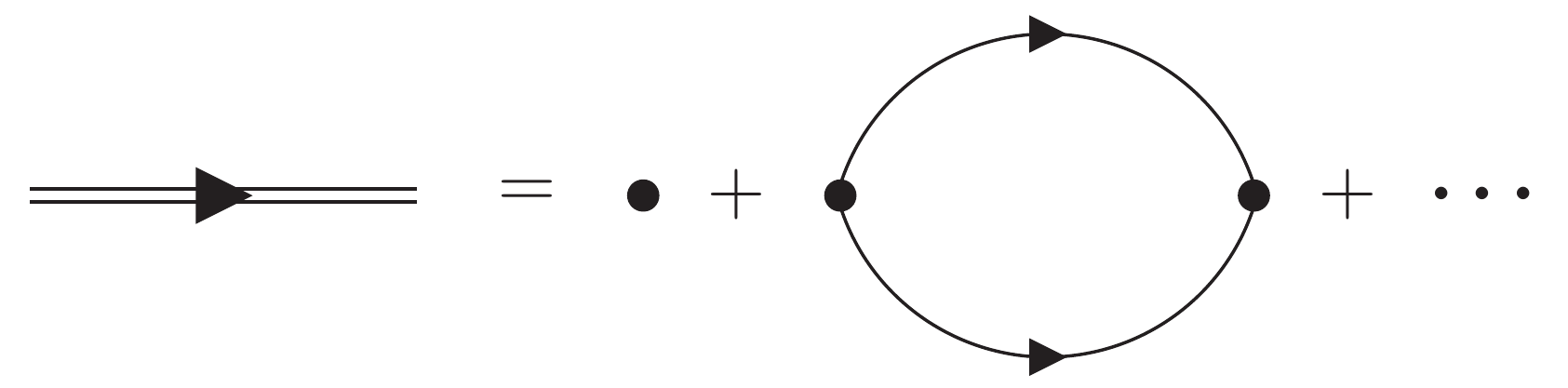}\\\bigskip
\includegraphics[width=0.4\columnwidth]{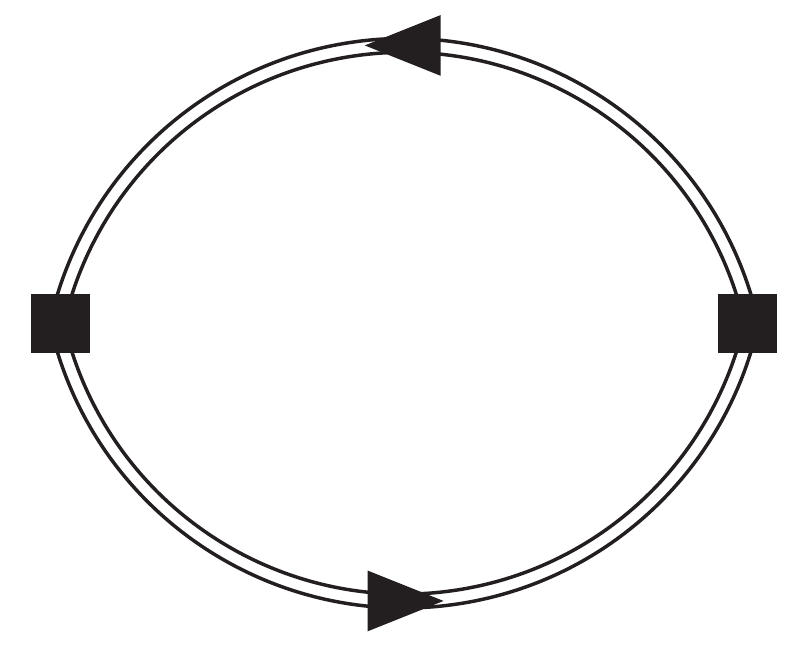}
\caption{\label{fig:diagram}
Diagrammatic representation of (top) the pair propagator and (bottom) the contact-contact response function, which become exact both in the dimer limit and in the high-temperature limit.
The single and double lines represent the fermion and pair propagators, respectively, whereas the dot is a bare coupling constant and the square is to insert the contact density operator.}
\end{figure}

In order to evaluate the contact-contact response function, we first introduce the pair propagator in the medium above the superfluid critical temperature~\cite{Melo:1993},
\begin{align}
[D(ip_0,\p)]^{-1} = \frac1g - \int_\q\,\frac{1 - f_F\!\left(\frac{(\p/2+\q)^2}{2m}\right)
- f_F\!\left(\frac{(\p/2-\q)^2}{2m}\right)}{ip_0-\frac{\p^2}{4m}-\frac{\q^2}{m}+2\mu},
\end{align}
whose diagrammatic representation is depicted in Fig.~\ref{fig:diagram}.
Here, $g<0$ is a bare coupling constant, $p_0=2\pi n/\beta$ is the bosonic Matsubara frequency, and $f_F(\eps)=1/[e^{\beta(\eps-\mu)}+1]$ is the Fermi-Dirac distribution function.
Figure~\ref{fig:diagram} also depicts the diagrammatic representation of the contact-contact response function,
\begin{align}
R_{\C\C}(ik_0,\k) = \frac{m^4}{\beta}\sum_{p_0}\int_\p\,D(ik_0+ip_0,\k+\p)D(ip_0,\p),
\end{align}
which fully incorporates two-body physics and thus becomes exact both in the dimer limit and in the high-temperature limit.
The summation over the bosonic Matsubara frequency can be performed by employing the complex contour integration together with the spectral representation of the pair propagator,
\begin{align}
D(ip_0,\p) = \int_{-\infty}^\infty\!\frac{dE}{\pi}\,\frac{\Im[D(E-i0^+,\p)]}{ip_0-E},
\end{align}
so that we obtain
\begin{align}
& R_{\C\C}(ik_0,\k) = -\iint_{-\infty}^\infty\!\frac{dE}{\pi}\frac{dE'}{\pi}\int_\p\,
\left(\frac1{e^{\beta E}-1}-\frac1{e^{\beta E'}-1}\right) \notag\\
&\quad \times \frac{\Im[m^2D(E-i0^+,\k+\p)]\Im[m^2D(E'-i0^+,\p)]}{ik_0+E-E'}.
\end{align}
Finally, by setting $ik_0\to w$ and $\k=\0$ and changing the integration variables to $\eps^{(\prime)}=E^{(\prime)}-\frac{\p^2}{4m}+2\mu$, the contact-contact response function turns into
\begin{align}\label{eq:contact}
& R_{\C\C}(w,\0) \notag\\
&= -\iint_{-\infty}^\infty\!\frac{d\eps}{\pi}\frac{d\eps'}{\pi}\int_\p\,
\left[f_B\!\left(\eps+\tfrac{\p^2}{4m}\right) - f_B\!\left(\eps'+\tfrac{\p^2}{4m}\right)\right] \notag\\
&\quad \times \frac{\Im[m^2D_\p(\eps-i0^+)]\Im[m^2D_\p(\eps'-i0^+)]}{w+\eps-\eps'},
\end{align}
where $f_B(\eps)=1/[e^{\beta(\eps-2\mu)}-1]$ is the Bose-Einstein distribution function and $D_\p(\eps)\equiv D(\eps+\p^2/4m-2\mu,\p)$ is the pair propagator in the center-of-mass frame with the residual dependence on its momentum due to the medium.

Similarly, the contact density itself is provided by
\begin{align}
\C = -\frac{m^2}{\beta}\sum_{p_0}\int_\p\,e^{ip_00^+}D(ip_0,\p),
\end{align}
where the summation over the bosonic Matsubara frequency leads to
\begin{align}\label{eq:contact-density}
\C = \int_{-\infty}^\infty\!\frac{d\eps}{\pi}\int_\p\,
f_B\!\left(\eps+\tfrac{\p^2}{4m}\right)\Im[m^2D_\p(\eps-i0^+)]
\end{align}
with the same pair propagator introduced above.

\subsection{Dimer limit}\label{sec:dimer}
The pair propagator is further simplified in the dimer limit, $a\to+0$, where $f_F(\eps>0)$ and $f_B(\eps>0)$ are negligible because of $2\mu=-1/ma^2\to-\infty$, so that the pair propagator is reduced to that in the vacuum.
The integration over $\q$ can thus be performed to lead to
\begin{align}\label{eq:vacuum}
D(\eps-i0^+) = \frac{\Omega_{d-1}}{m}\frac{d-2}{a^{2-d} - (-m\eps+i0^+)^{d/2-1}},
\end{align}
where the subscript of $\p$ is dropped on the left-hand side and the bare coupling constant is replaced by
\begin{align}\label{eq:coupling}
g = \frac{\Omega_{d-1}}{m}\frac{d-2}{a^{2-d}-\frac{\Lambda^{d-2}}{\Gamma(d/2)\Gamma(2-d/2)}}
\end{align}
in the cutoff regularization with $|\q|<\Lambda$.
By substituting its imaginary part,
\begin{align}\label{eq:imaginary}
\Im[D(\eps-i0^+)] &= \theta(a)\,\frac{2\pi\Omega_{d-1}}{m^2a^{4-d}}\,
\delta\!\left(\eps+\tfrac1{ma^2}\right) \notag\\
&\quad + \theta(\eps)\Im[D(\eps-i0^+)],
\end{align}
into the contact-contact response function in Eq.~(\ref{eq:contact}), we obtain
\begin{align}\label{eq:contact_dimer}
R_{\C\C}(w,\0) &= \frac{\Omega_{d-1}m^2\N}{a^{4-d}}
\int_0^\infty\!\frac{d\eps}{\pi}\Im[D(\eps-i0^+)] \notag\\
&\quad \times \left(\frac1{\eps+\frac1{ma^2}-w} + \frac1{\eps+\frac1{ma^2}+w}\right),
\end{align}
where $\N=2\int_\p f_B(\p^2/4m-1/ma^2)$ is the number density with $1/ma^2$ being the binding energy of dimers.

\begin{figure}[t]
\includegraphics[width=\columnwidth]{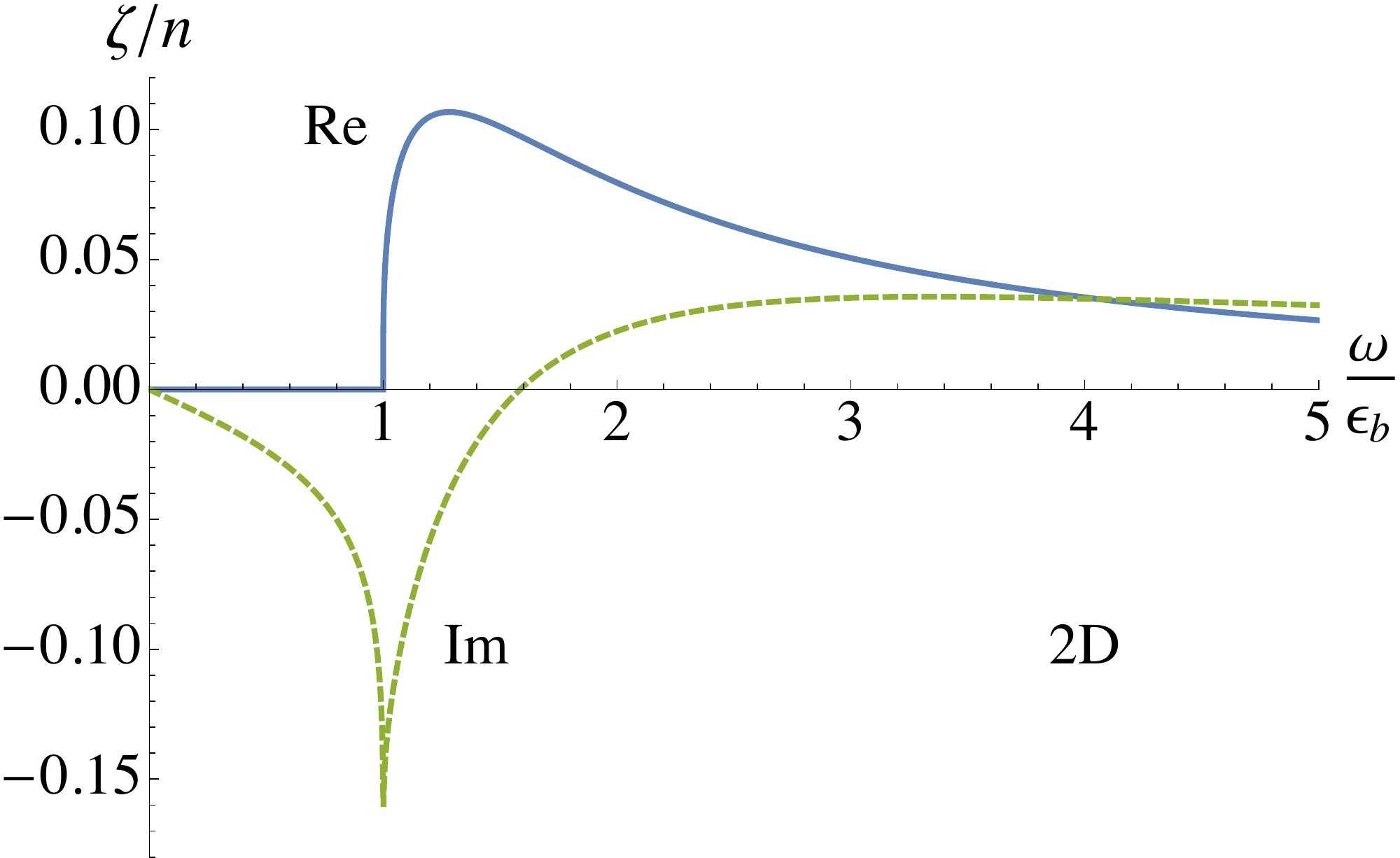}\\\bigskip
\includegraphics[width=\columnwidth]{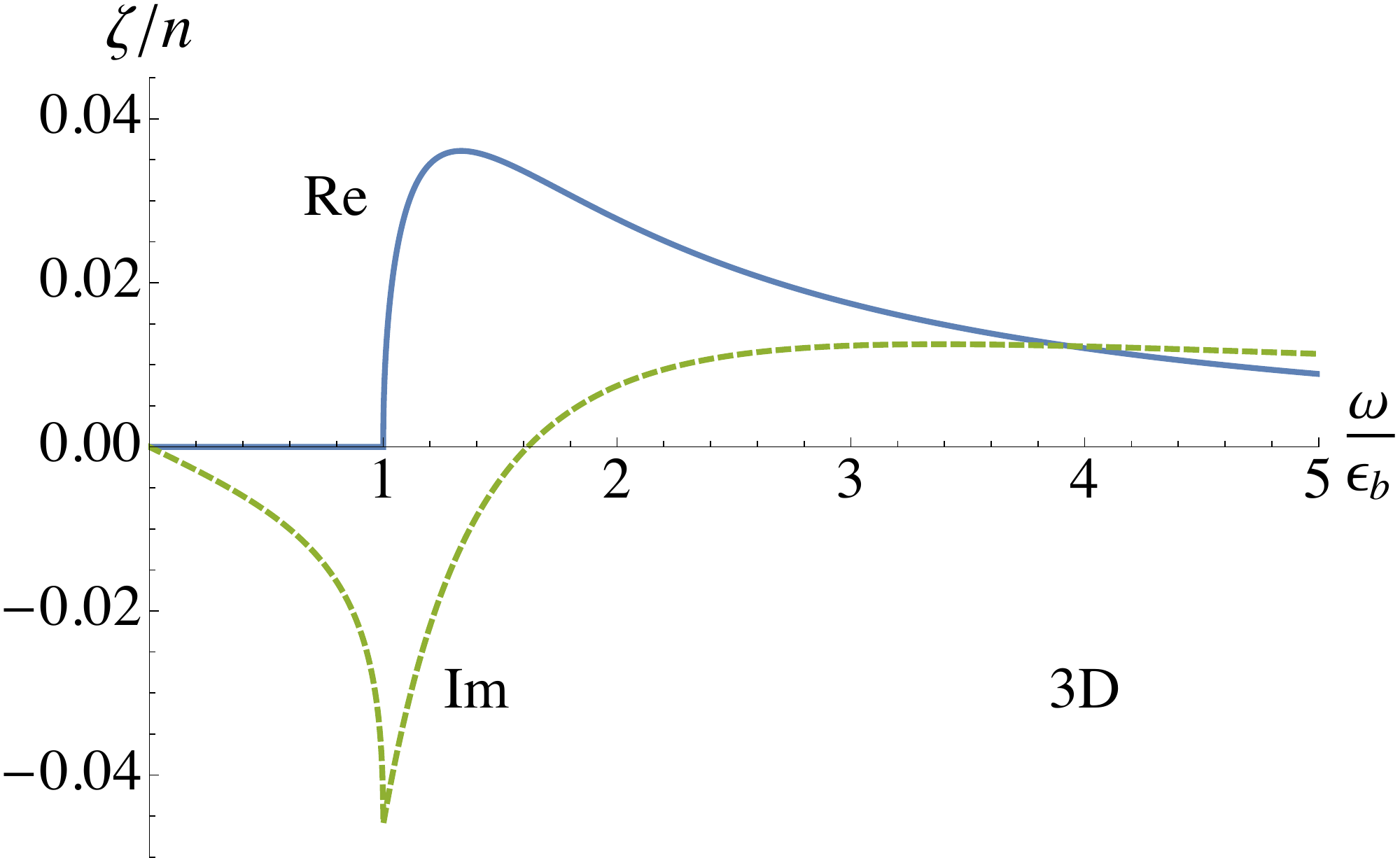}
\caption{\label{fig:bulk}
Frequency-dependent complex bulk viscosity $\zeta(\omega+i0^+)$ in the dimer limit for (top) $d=2$ and (bottom) $d=3$ in units of the number density.
Its real and imaginary parts are plotted by the solid and dashed curves, respectively, and the frequency is normalized by the binding energy of dimers.}
\end{figure}

Now, turning to thermodynamics in the dimer limit, the pressure of a free Bose gas of tightly bound dimers is provided by
\begin{align}
p = -\frac1\beta\int_\p\,\ln\!\left\{1
- \exp\!\left[-\beta\left(\tfrac{\p^2}{4m}-\tfrac1{ma^2}-2\mu\right)\right]\right\},
\end{align}
from which all thermodynamic variables are readily obtained including
\begin{align}
\left(\frac{\d\C}{\d a}\right)_{\N,\S} = -\frac{\Omega_{d-1}(4-d)\N}{a^{5-d}}.
\end{align}
Then, by employing the following identity,%
\footnote{This follows by taking the limit of $w\to-1/ma^2$ on both sides of the spectral representation, $D(w)=\int_{-\infty}^\infty\!d\eps\,\Im[D(\eps-i0^+)]/[\pi(w-\eps)]$, together with Eq.~(\ref{eq:imaginary}).}
\begin{align}
\frac{\Omega_{d-1}(4-d)}{2ma^{2-d}} = \int_0^\infty\!\frac{d\eps}{\pi}\,
\frac{\Im[D(\eps-i0^+)]}{\eps+\frac1{ma^2}},
\end{align}
and comparing it with Eq.~(\ref{eq:contact_dimer}), the sum rule is found to be related to the contact-contact response function at $w=0$ as
\begin{align}\label{eq:sum-rule_dimer}
-\frac{a^{3-d}}{d^2\cdot\Omega_{d-1}m}\left(\frac{\d\C}{\d a}\right)_{\N,\S}
= \frac{R_{\C\C}(0,\0)}{(d\cdot\Omega_{d-1}ma^{d-2})^2}.
\end{align}
Accordingly, the substitution of Eqs.~(\ref{eq:contact_dimer}) and (\ref{eq:sum-rule_dimer}) into Eq.~(\ref{eq:bulk_zero-range}) leads to
\begin{align}\label{eq:bulk_dimer}
\zeta(w) &= \frac\N{d^2\cdot\Omega_{d-1}a^d}
\int_0^\infty\!\frac{d\eps}{i\pi}\,\frac{\Im[D(\eps-i0^+)]}{\eps+\frac1{ma^2}} \notag\\
&\quad \times \left(\frac1{\eps+\frac1{ma^2}-w} - \frac1{\eps+\frac1{ma^2}+w}\right).
\end{align}
The resulting frequency-dependent complex bulk viscosity is plotted in Fig.~\ref{fig:bulk} for $w\to\omega+i0^+$, which is exact in the limit of $a\to+0$ at fixed temperature and number density.
In particular, we find that its real part indeed vanishes at a lower frequency than the binding energy of dimers without the unphysical divergence at zero frequency,%
\footnote{\label{footnote}
We note that, if the contact-contact response function in Eq.~(\ref{eq:contact}) was substituted into the bulk viscosity formula of Eq.~(\ref{eq:kubo_contact}), it would give rise to a divergent term of $\Re[\zeta(\omega+i0^+)]=(\d\N/\d\mu)_\beta\,\pi\delta(\omega)/(d\cdot ma^2)^2+\cdots$ in the dimer limit.}
whereas the bound-continuum transition turns possible above the dimer-breakup threshold.
We also note that our $\Re[\zeta(\omega+i0^+)]$ in the dimer limit for $d=2$ coincides with the zero-temperature and zero-density limit of the bulk viscosity spectral function in Ref.~\cite{Taylor:2012}.

\section{High-temperature limit}\label{sec:high-T}
The diagrammatic method employed in the previous section is also applicable to the high-temperature limit, where our Kubo formalism can be contrasted with the kinetic theory.

\subsection{Quantum virial expansion}
The quantum virial expansion is a systematic expansion in terms of fugacity, $z=e^{\beta\mu}$, which becomes small in the high-temperature limit at fixed number density and scattering length~\cite{Liu:2013}.
Because of $f_F(\eps)\to ze^{-\beta\eps}$ and $f_B(\eps)\to z^2e^{-\beta\eps}$ to the lowest order in fugacity, Eq.~(\ref{eq:contact}) after the integration over $\p$ is reduced to
\begin{align}\label{eq:contact_high-T}
& R_{\C\C}(w,\0) = -\frac{2^{d/2}z^2}{\lambda_T^d}
\iint_{-\infty}^\infty\!\frac{d\eps}{\pi}\frac{d\eps'}{\pi}
\frac{e^{-\beta\eps} - e^{-\beta\eps'}}{w+\eps-\eps'} \notag\\
&\times \Im[m^2D(\eps-i0^+)]\Im[m^2D(\eps'-i0^+)] + O(z^3),
\end{align}
where $\lambda_T=\sqrt{2\pi\beta/m}$ is the thermal de Broglie wavelength and $D(\eps-i0^+)$ provided by Eq.~(\ref{eq:vacuum}) is the pair propagator in the vacuum.
The resulting contact-contact response function indeed reproduces Eq.~(39) of Ref.~\cite{Nishida:2019} derived systematically with a different method.

Similarly, the contact density in Eq.~(\ref{eq:contact-density}) is reduced to
\begin{align}
\C = \frac{2^{d/2}z^2}{\lambda_T^d}\int_{-\infty}^\infty\!\frac{d\eps}{\pi}\,
e^{-\beta\eps}\Im[m^2D(\eps-i0^+)] + O(z^3).
\end{align}
Because its partial derivative with respect to $a$ at fixed $\N$ and $\S$ is equivalent to that at fixed $\beta$ and $z$ to the lowest order in fugacity, we obtain
\begin{align}
& \left(\frac{\d\C}{\d a}\right)_{\N,\S}
= \frac{2^{d/2}z^2}{\lambda_T^d}\frac{a^{1-d}}{\Omega_{d-1}m}
\iint_{-\infty}^\infty\!\frac{d\eps}{\pi}\frac{d\eps'}{\pi}
\frac{e^{-\beta\eps} - e^{-\beta\eps'}}{\eps-\eps'} \notag\\
&\times \Im[m^2D(\eps-i0^+)]\Im[m^2D(\eps'-i0^+)] + O(z^3),
\end{align}
where Eqs.~(42)--(44) of Ref.~\cite{Nishida:2019} are followed in reverse.
Then, by comparing it with Eq.~(\ref{eq:contact_high-T}), the sum rule is found to be related to the contact-contact response function at $w=0$ in the same way as Eq.~(\ref{eq:sum-rule_dimer}).
Accordingly, the substitution of Eqs.~(\ref{eq:contact_high-T}) and (\ref{eq:sum-rule_dimer}) into Eq.~(\ref{eq:bulk_zero-range}) leads to
\begin{align}\label{eq:bulk_high-T}
& \zeta(w) = \frac{2^{d/2}z^2}{(d\cdot\Omega_{d-1}a^{d-2})^2\lambda_T^d}
\iint_{-\infty}^\infty\!\frac{d\eps}{\pi}\frac{d\eps'}{\pi}
\frac{e^{-\beta\eps} - e^{-\beta\eps'}}{\eps-\eps'} \notag\\
&\times \frac{\Im[mD(\eps-i0^+)]\Im[mD(\eps'-i0^+)]}{i\,(w+\eps-\eps')} + O(z^3).
\end{align}
Therefore, we find that the real part of the frequency-dependent complex bulk viscosity for $w\to\omega+i0^+$ reproduces the bulk viscosity spectral function in Refs.~\cite{Nishida:2019,Enss:2019,Hofmann:2020}, cf.\ Eq.~(40) of Ref.~\cite{Nishida:2019}.
In particular, it gives rise to a term proportional to $\delta(\omega)/a^4$ for $a>0$ originating from the bound-bound transition.
As opposed to the dimer limit, such a zero-frequency peak at $O(z^2)$ is physical in the high-temperature limit and to be broadened by resumming higher-order corrections in fugacity, for example, due to atom-dimer and dimer-dimer collisions [see the Appendix of Ref.~\cite{Nishida:2019} for the $O(z^3)$ correction].
How to systematically resum such higher-order corrections is currently unknown and needs to be elucidated in a future study.

\begin{figure}[t]
\includegraphics[width=0.8\columnwidth]{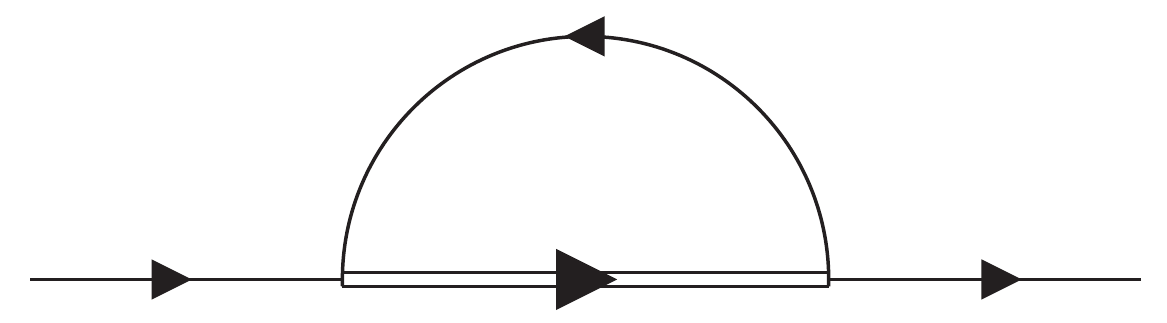}
\caption{\label{fig:self-energy}
Diagrammatic representation of the fermion self-energy.
See the caption of Fig.~\ref{fig:diagram} for the other details.}
\end{figure}

For later purpose, we also evaluate the fermion self-energy,
\begin{align}
\Sigma(ip'_0,\p) = \frac1\beta\sum_{q'_0}\int_\q\,D(ip'_0+iq'_0,\p+\q)G(iq'_0,\q),
\end{align}
whose diagrammatic representation is depicted in Fig.~\ref{fig:self-energy}.
Here, $G(iq'_0,\q)=1/(iq'_0-\q^2/2m+\mu)$ with $q'_0=2\pi(n+1/2)/\beta$ is the fermion propagator and the summation over the fermionic Matsubara frequency leads to
\begin{align}
\Sigma(ip'_0,\p) &= \int_{-\infty}^\infty\!\frac{d\eps}{\pi}\int_\q
\left[f_B\!\left(\eps+\tfrac{(\p+\q)^2}{4m}\right)
+ f_F\!\left(\tfrac{\q^2}{2m}\right)\right] \notag\\
&\quad \times \frac{\Im[D_{\p+\q}(\eps-i0^+)]}
{ip'_0+\mu-\eps-\frac{(\p+\q)^2}{4m}+\frac{\q^2}{2m}}.
\end{align}
Because of $f_F(\eps)\to ze^{-\beta\eps}$ and $f_B(\eps)\to z^2e^{-\beta\eps}$ to the lowest order in fugacity, the first term in the square brackets is negligible and the integration over $\eps$ can thus be performed, so that we obtain
\begin{align}\label{eq:self-energy}
\Sigma(ip'_0,\p) &= \int_\q\,f_F\!\left(\tfrac{\q^2}{2m}\right)
D\!\left(ip'_0-\tfrac{\p^2}{2m}+\mu+\tfrac{(\p-\q)^2}{4m}\right) \notag\\
&\quad + O(z^2),
\end{align}
which has both real and imaginary parts at $O(z)$~\cite{Dusling:2013,Chafin:2013}.

The momentum distribution function of fermions for each spin component then follows from
\begin{align}
f_\p &= \frac1\beta\sum_{p'_0}e^{ip'_00^+}[G(ip'_0,\p) \notag\\
&\quad + G(ip'_0,\p)\Sigma(ip'_0,\p)G(ip'_0,\p) + \cdots],
\end{align}
where the summation over the fermionic Matsubara frequency leads to
\begin{align}
f_\p &= f_F\!\left(\tfrac{\p^2}{2m}\right)
+ \int_{-\infty}^\infty\!\frac{d\eps}{\pi}\,f_F\!\left(\eps+\tfrac{\p^2}{2m}\right) \notag\\
&\quad \times \Im\!\left[\frac{\Sigma\bigl(\eps+\frac{\p^2}{2m}-\mu-i0^+,\p\bigr)}
{(\eps-i0^+)^2}\right] + O(z^3).
\end{align}
Equivalently, it can also be expressed as
\begin{align}\label{eq:distribution}
f_\p &= Z_\p\,f_F\!\left(\tfrac{\p^2}{2m}\right)
+ f'_F\!\left(\tfrac{\p^2}{2m}\right)
\Re\!\left[\Sigma\!\left(\tfrac{\p^2}{2m}-\mu-i0^+,\p\right)\right] \notag\\
&\quad + \mathrm{P}\int_{-\infty}^\infty\!\frac{d\eps}{\pi}\,\frac1\eps\,
\frac\d{\d\eps}\Bigl\{f_F\!\left(\eps+\tfrac{\p^2}{2m}\right) \notag\\
&\qquad \times 
\Im\!\left[\Sigma\!\left(\eps+\tfrac{\p^2}{2m}-\mu-i0^+,\p\right)\right]\Bigr\} + O(z^3),
\end{align}
where $Z_\p=1+\Re[\Sigma'(\p^2/2m-\mu-i0^+,\p)]$ is the quasiparticle residue and $'$ denotes the partial derivative with respect to $\eps$ at $\eps=0$.
In particular, the last term involving the imaginary part of the self-energy is responsible for the characteristic large-momentum tail of $\lim_{|\p|\to\infty}f_\p=\C/|\p|^4$ determined by the contact density~\cite{Tan:2008a,Tan:2008b,Tan:2008c}.

\subsection{Kinetic theory}
The bulk viscosity is provided by $\zeta=\lim_{w\to i0^+}\zeta(w)$, which at $a\to\infty$ following from Eq.~(\ref{eq:bulk_high-T}) in the high-temperature limit was found to disagree with that derived from the kinetic theory both for $d=2,\,3$~\cite{Nishida:2019,Enss:2019,Hofmann:2020}.
Here, we discuss that such discrepancies between the Kubo formalism and the kinetic theory for the bulk viscosity are attributed to the fact that the quasiparticle approximation assumed by the latter breaks down even in the high-temperature limit where the fermion self-energy becomes small.

The bulk viscosity in the high-temperature limit was computed in Refs.~\cite{Dusling:2013,Chafin:2013} by employing the Landau kinetic equation for quasiparticles,
\begin{align}\label{eq:kinetic}
\frac{\d f_\p}{\d t} + \frac{\d E_\p}{\d\p}\cdot\frac{\d f_\p}{\d\r}
- \frac{\d E_\p}{\d\r}\cdot\frac{\d f_\p}{\d\p}
= \left(\frac{\d f_\p}{\d t}\right)_\mathrm{coll},
\end{align}
where
\begin{align}\label{eq:quasiparticle}
& E_\p[f(t,\r)] = \frac{\p^2}{2m} \notag\\
&\quad + \Re\!\left[\Sigma\!\left(\tfrac{\p^2}{2m}-\mu-i0^+,\p\right)\right]
\Big|_{f_F(\q^2/2m)\to f_\q(t,\r)}
\end{align}
is the quasiparticle energy functional of the nonequilibrium distribution function and its on-shell self-energy correction is obtained from the real part of Eq.~(\ref{eq:self-energy}) with the Fermi-Dirac distribution function replaced by the nonequilibrium distribution function.
In particular, the scale invariance breaking in the quasiparticle energy due to its self-energy correction was found to be essential to a nonvanishing bulk viscosity~\cite{Dusling:2013,Chafin:2013}.
However, we consider that such a kinetic equation is not fully grounded because the self-energy in Eq.~(\ref{eq:self-energy}) has both real and imaginary parts at $O(z)$.
Namely, if the real part of the self-energy is essential to the bulk viscosity, its imaginary part being at the same order in fugacity is non-negligible so as to invalidate the quasiparticle approximation, i.e., replacing the fermion spectral function by a $\delta$ function on which the kinetic equation is founded~\cite{Kita:2010}.%
\footnote{It should be emphasized that our argument herein does not apply to the Boltzmann equation to compute the shear viscosity in the high-temperature limit because both real and imaginary parts of the self-energy are consistently neglected~\cite{Massignan:2005,Bruun:2005,Bruun:2012,Schafer:2012}.
Although such a Boltzmann equation merely leads to the vanishing bulk viscosity, it is indeed the correct ``leading'' behavior in the high-temperature limit.}

In order to further support our consideration, let us study the equilibrium distribution function resulting from the kinetic equation.
Because the collision term in Eq.~(\ref{eq:kinetic}) must be canceled under the conservation of quasiparticle energies~\cite{Baym-Pethick}, the equilibrium distribution function in the rest frame obeys the self-consistent equation of
\begin{align}
f_\p = \frac1{e^{\beta(E_\p[f]-\mu)}+1}.
\end{align} 
By substituting the quasiparticle energy in Eq.~(\ref{eq:quasiparticle}) and expanding the right-hand side in terms of fugacity iteratively, we obtain
\begin{align}
f_\p &= f_F\!\left(\tfrac{\p^2}{2m}\right) + f'_F\!\left(\tfrac{\p^2}{2m}\right)
\Re\!\left[\Sigma\!\left(\tfrac{\p^2}{2m}-\mu-i0^+,\p\right)\right] \notag\\
&\quad + O(z^3),
\end{align}
where two contributions are found to be missing from the microscopic distribution function in Eq.~(\ref{eq:distribution}).
One is the factor of the quasiparticle residue, whereas the other is the whole term involving the imaginary part of the self-energy.%
\footnote{These two are actually related because the quasiparticle residue originating from the frequency dependence in the real part of the self-energy necessarily leads to the presence of the imaginary part according to the Kramers-Kronig relations.}
Furthermore, because all thermodynamic variables in the kinetic theory are expressed in terms of the distribution function~\cite{Baym-Pethick}, they also differ from the microscopic ones in the quantum virial expansion.
Therefore, the Landau kinetic equation employed in Refs.~\cite{Dusling:2013,Chafin:2013} is incapable of describing physics at the order where the self-energy contributes because its imaginary part neglected therein is actually non-negligible.
We consider that this constitutes the origin of the discrepancy between the Kubo formalism and the kinetic theory for the bulk viscosity.

\section{Summary}\label{sec:summary}
The standard Kubo formula for the bulk viscosity presented in Eq.~(\ref{eq:kubo_stress}) has flaws to give rise to unphysical divergences at zero frequency.
They are however resolved with the corrected Kubo formula~\cite{Mori:1962,Luttinger:1964,Bradlyn:2012}, which has been known since long ago but is not well appreciated by the literature in the context of ultracold atom physics.
After carefully rederiving the Kubo formula for the frequency-dependent complex bulk viscosity as well as its sum rule, we found that the sum rule for two-component fermions with a zero-range interaction in two and three dimensions [Eq.~(\ref{eq:sum-rule_zero-range})] coincides with that derived in Ref.~\cite{Taylor:2010}, although we do not fully agree with the derivation therein because of the neglected pressure fluctuations.

The Kubo formula can be evaluated unambiguously, in particular, by working with the complex bulk viscosity on the upper-half plane of complex frequency.
We then confirmed that the bulk viscosity spectral function indeed vanishes at an arbitrary frequency in the free and unitarity limits and at a lower frequency than the binding energy of dimers in the dimer limit [Eq.~(\ref{eq:bulk_dimer})] without the unphysical divergences at zero frequency.

In the high-temperature limit, the bulk viscosity spectral function in the quantum virial expansion [Eq.~(\ref{eq:bulk_high-T})] was reproduced with our diagrammatic method.
We also discussed that the Landau kinetic equation employed in Refs.~\cite{Dusling:2013,Chafin:2013} to compute the bulk viscosity is not fully grounded even in the high-temperature limit where the fermion self-energy becomes small.
This is because the self-energy has both real and imaginary parts at the same order in fugacity so as to invalidate the quasiparticle approximation, i.e., replacing the fermion spectral function by a $\delta$ function on which the kinetic equation is founded.
We consider that this constitutes the origin of the recently found discrepancy between the Kubo formalism and the kinetic theory for the bulk viscosity~\cite{Nishida:2019,Enss:2019,Hofmann:2020}.

\acknowledgments
The authors thank Tilman Enss, Yoshimasa Hidaka, and Masaru Hongo for valuable discussions.
This work was supported by JSPS KAKENHI Grants No.\ JP15K17727, No.\ JP15H05855, and No.\ JP19J13698.
One of the authors (K.F.) also benefited from the RIKEN iTHEMS Program as a student trainee.

\newpage\onecolumngrid\appendix
\section{Kubo's canonical correlation function}\label{sec:canonical}
In this Appendix, we present some useful formulas and their detailed derivations regarding Kubo's canonical correlation function~\cite{Kubo:1957}.

\subsection{Derivation of Eq.~(\ref{eq:current-stress})}\label{sec:current-stress}
First, by multiplying the response function defined in Eq.~(\ref{eq:response}) by $iw$ and using $iwe^{iwt}=\d_te^{iwt}$, the temporal integration by parts leads to
\begin{align}
iwR_{\J_i\J_j}(w,\k)
= -i\int\!d\r\,e^{-i\k\cdot\r}\<[\hat\J_i(0,\r),\hat\J_j(0,\0)]\>
- i\int_0^\infty\!dt\int\!d\r\,e^{iwt-i\k\cdot\r}\<[\d_t\hat\J_i(t,\r),\hat\J_j(0,\0)]\>.
\end{align}
Here, the first term turns out to vanish because of $[\hat\J_i(\r),\hat\J_j(\r')]=[\hat\J_j(\r)\d_i+\hat\J_i(\r')\d_j]\delta(\r-\r')/(im)$~\cite{Nishida:2007}.
Then, by using the momentum continuity equation~(\ref{eq:continuity}), the spatial integration by parts leads to
\begin{align}
iwR_{\J_i\J_j}(w,\k)
= -\frac{k_k}{m}\int_0^\infty\!dt\int\!d\r\,e^{iwt-i\k\cdot\r}
\<[\hat\pi_{ik}(t,\r),\hat\J_j(0,\0)]\>.
\end{align}
After using the spacetime translational invariance, the integrand can be rewritten as
\begin{align}
\<[\hat\pi_{ik}(0,\0),\hat\J_j(-t,-\r)]\>
&= \<[\hat\pi_{ik}(0,\0)\hat\J_j(-t,-\r)
- \hat\pi_{ik}(0,\0)e^{-\beta\hat{H}}\hat\J_j(-t,-\r)e^{\beta\hat{H}}]\> \notag\\
&= -\int_0^\beta\!d\tau\,\d_\tau\<\hat\pi_{ik}(0,\0)e^{-\tau\hat{H}}\hat\J_j(-t,-\r)e^{\tau\hat{H}}\> \notag\\
&= i\int_0^\beta\!d\tau\,\<\hat\pi_{ik}(0,\0)\d_t\hat\J_j(-t+i\tau,-\r)\>,
\end{align}
because the number operator $\hat{N}$ commutes with the other operators.
Then, by using the momentum continuity equation~(\ref{eq:continuity}) again, the spatial integration by parts leads to
\begin{align}
iwR_{\J_i\J_j}(w,\k)
= -\frac{k_kk_l}{m^2}\int_0^\infty\!dt\int\!d\r\,e^{iwt-i\k\cdot\r}\int_0^\beta\!d\tau\,
\<\delta\hat\pi_{ik}(0,\0)\delta\hat\pi_{jl}(-t+i\tau,-\r)\>,
\end{align}
where the expectation value needs to be subtracted from the operator to ensure that boundary contributions at spatial infinity vanish under the clustering property:
\begin{align}
\lim_{|\r|\to\infty}\<\delta\hat\pi_{ik}(0,\0)\delta\hat\pi_{jl}(-t+i\tau,-\r)\>
= \lim_{|\r|\to\infty}\<\delta\hat\pi_{ik}(0,\0)\>\<\delta\hat\pi_{jl}(-t+i\tau,-\r)\> = 0.
\end{align}
Finally, by using the spacetime translational invariance again and comparing the outcome with the canonical correlation function defined in Eq.~(\ref{eq:canonical}), we arrive at Eq.~(\ref{eq:current-stress}).

\subsection{Derivation of Eq.~(\ref{eq:bulk_kubo})}\label{sec:fluctuation}
First, by substituting the modified stress tensor operator defined in Eq.~(\ref{eq:fluctuation}) into the right-hand side of Eq.~(\ref{eq:bulk_kubo}), we obtain
\begin{align}
K_{\tilde\pi\tilde\pi}(w,\0) &= K_{\pi\pi}(w,\0)
- 2\left(\frac{\d p}{\d\N}\right)_\E K_{\N\pi}(w,\0)
- 2\left(\frac{\d p}{\d\E}\right)_\N K_{\H\pi}(w,\0) \notag\\
&\quad + \left(\frac{\d p}{\d\N}\right)_\E
\left[\left(\frac{\d p}{\d\N}\right)_\E K_{\N\N}(w,\0)
+ \left(\frac{\d p}{\d\E}\right)_\N K_{\N\H}(w,\0)\right] \notag\\
&\quad + \left(\frac{\d p}{\d\E}\right)_\N
\left[\left(\frac{\d p}{\d\N}\right)_\E K_{\H\N}(w,\0)
+ \left(\frac{\d p}{\d\E}\right)_\N K_{\H\H}(w,\0)\right],
\end{align}
because $\hat{N}$ and $\hat{H}$ are conserved.
Then, by using the following properties of the canonical correlation function,
\begin{align}
K_{\N\O}(w,\0) &= -\frac1{iw}\left(\frac{\d\<\hat\O(\0)\>}{\beta\d\mu}\right)_\beta, \\
K_{\H\O}(w,\0) &= \frac1{iw}\left(\frac{\d\<\hat\O(\0)\>}{\d\beta}\right)_{\beta\mu},
\end{align}
for $\<\hat\N(\r)\>=\N$, $\<\hat\H(\r)\>=\E$, and $\<\hat\pi(\r)\>=p$, the thermodynamic identities lead to
\begin{align}
K_{\tilde\pi\tilde\pi}(w,\0) &= K_{\pi\pi}(w,\0)
+ \frac2{iw}\left(\frac{\d p}{\d\N}\right)_\E\left(\frac{\d p}{\beta\d\mu}\right)_\beta
- \frac2{iw}\left(\frac{\d p}{\d\E}\right)_\N\left(\frac{\d p}{\d\beta}\right)_{\beta\mu} \notag\\
&\quad - \frac1{iw}\left(\frac{\d p}{\d\N}\right)_\E
\left[\left(\frac{\d p}{\d\N}\right)_\E\left(\frac{\d\N}{\beta\d\mu}\right)_\beta
+ \left(\frac{\d p}{\d\E}\right)_\N\left(\frac{\d\E}{\beta\d\mu}\right)_\beta\right] \notag\\
&\quad + \frac1{iw}\left(\frac{\d p}{\d\E}\right)_\N
\left[\left(\frac{\d p}{\d\N}\right)_\E\left(\frac{\d\N}{\d\beta}\right)_{\beta\mu}
+ \left(\frac{\d p}{\d\E}\right)_\N\left(\frac{\d\E}{\d\beta}\right)_{\beta\mu}\right] \notag\\
&= K_{\pi\pi}(w,\0) + \frac1{iw}\left(\frac{\d p}{\d\N}\right)_\E\frac\N\beta
+ \frac1{iw}\left(\frac{\d p}{\d\E}\right)_\N\frac{\E+p}{\beta}.
\end{align}
Finally, by using the sound velocity in Eq.~(\ref{eq:sound}) and comparing the outcome with the frequency-dependent complex bulk viscosity defined in Eq.~(\ref{eq:bulk_dynamic}), we arrive at Eq.~(\ref{eq:bulk_kubo}).

\subsection{Comparison of Eq.~(\ref{eq:bulk_dynamic}) with Ref.~\cite{Bradlyn:2012}}\label{sec:comparison}
First, by using $e^{iwt}=\d_te^{iwt}/iw$ in the canonical correlation function in Eq.~(\ref{eq:bulk_dynamic}), the temporal integration by parts leads to
\begin{align}
\zeta(w) = \frac{mc_s^2\N}{iw} - \frac1{iw}\int\!d\r\int_0^\beta\!d\tau\,
\<\delta\hat\pi(-i\tau,\r)\delta\hat\pi(0,\0)\>
- \frac1{iw}\int_0^\infty\!dt\int\!d\r\,e^{iwt}\int_0^\beta\!d\tau\,
\<\d_t\delta\hat\pi(t-i\tau,\r)\delta\hat\pi(0,\0)\>.
\end{align}
Then, the integral over $\tau$ in the last term can be rewritten as
\begin{align}
\int_0^\beta\!d\tau\,\<\d_t\delta\hat\pi(t-i\tau,\r)\delta\hat\pi(0,\0)\>
&= i\int_0^\beta\!d\tau\,\d_\tau
\<e^{\tau\hat{H}}\delta\hat\pi(t,\r)e^{-\tau\hat{H}}\delta\hat\pi(0,\0)\> \notag \\
&= i\<[e^{\beta\hat{H}}\delta\hat\pi(t,\r)e^{-\beta\hat{H}}\delta\hat\pi(0,\0)
- \delta\hat\pi(t,\r)\delta\hat\pi(0,\0)]\> \notag \\
&= -i\<[\delta\hat\pi(t,\r),\delta\hat\pi(0,\0)]\> \notag \\
&= -i\<[\hat\pi(t,\r),\hat\pi(0,\0)]\>,
\end{align}
so that we obtain
\begin{align}
\zeta(w) = \frac{mc_s^2\N}{iw}
- \frac1{iw}\int\!d\r\int_0^\beta\!d\tau\,\<\delta\hat\pi(-i\tau,\r)\delta\hat\pi(0,\0)\>
+ \frac{R_{\pi\pi}(w,\0)}{iw}.
\end{align}
Here, the last term is the stress-stress response function, whereas the first and second terms evidently correspond to the inverse compressibility and equal-time commutator (``contact'') terms of Ref.~\cite{Bradlyn:2012}, respectively.

Actually, the first and second terms are combined into the sum rule in Eq.~(\ref{eq:sum-rule}), so that the frequency-dependent complex bulk viscosity can also be expressed as
\begin{align}
\zeta(w) = \frac{R_{\pi\pi}(w,\0)}{iw}
- \frac1{iw}\int_{-\infty}^\infty\!\frac{d\omega}{\pi}\,\zeta(\omega+i0^+).
\end{align}
This is the formula employed in Sec.~\ref{sec:contact}.
Its real part for $w\to\omega+i0^+$ then reads
\begin{align}
\Re[\zeta(\omega+i0^+)] = \frac{\Im[R_{\pi\pi}(\omega+i0^+,\0)]}{\omega}
- \pi\delta(\omega)\left[\Re[R_{\pi\pi}(i0^+,\0)]
- \int_{-\infty}^\infty\!\frac{d\omega'}{\pi}\zeta(\omega'+i0^+)\right],
\end{align}
whose last term unless canceled is missing from the standard Kubo formula for the bulk viscosity in Eq.~(\ref{eq:kubo_stress}).

\twocolumngrid

\end{document}